\documentclass[english]{article}
\usepackage[T1]{fontenc}
\usepackage[utf8]{luainputenc}
\usepackage{geometry}
\usepackage{float}
\usepackage{textcomp}
\usepackage{tipa}
\usepackage{amsmath}
\usepackage{amsthm}
\usepackage{amssymb}
\usepackage{stmaryrd}
\usepackage{graphicx}

\makeatletter

\newcommand*\LyXThinSpace{\,\hspace{0pt}}
\newcommand*\LyXHairSpace{\hspace{1pt}}
\providecommand{\tabularnewline}{\\}

\theoremstyle{definition}
\newtheorem{defn}{\protect\definitionname}
\theoremstyle{plain}
\newtheorem{thm}{\protect\theoremname}

\makeatother

\usepackage{babel}
\usepackage{listings}
\providecommand{\definitionname}{Definition}
\providecommand{\theoremname}{Theorem}

\begin{document}
\title{The Information Content and Entropy of Finite Patterns from a Combinatorial
Perspective}
\author{Zsolt Pocze}
\maketitle
\begin{abstract}
A unified combinatorial definition of the information content and
entropy of different types of patterns, compatible with the traditional
concepts of information and entropy, going beyond the limitations
of Shannon information interpretable for ergodic Markov processes.
We compare the information content of various finite patterns and
derive general properties of information quantity from these comparisons.
Using these properties, we define normalized information estimation
methods based on compression algorithms and Kolmogorov complexity.
From a combinatorial point of view, we redefine the concept of entropy
in a way that is asymptotically compatible with traditional entropy.
\end{abstract}

\section{Introduction}

The characteristic feature of a material is its pattern, which we
interpret broadly as the arrangement of its elementary components.
This also includes the relationships between individual parts. In
physical reality, every finite-dimensional pattern can be modeled,
with some accuracy, as a one-dimensional finite sequence. Therefore,
we examine information and entropy in relation to finite sequences
(patterns). Let $X^{*}$denote the set of finite patterns, where $X$
is the value set (or base set) of the patterns. For a given pattern
$A\in X^{*}$, $A=(x_{1},x_{2},...,x_{n})$, let $n=|A|$ be the length
of the pattern, $k=|X|$ be the number of elements in the value set,
and let $f(x)$, $x\in X$ be the number of occurrences of the element
$x$ in the pattern.

Information is nothing other than the number of binary decisions \cite{SHANNON_MTC},
required to uniquely determine a pattern, i.e., the number of decisions
needed to select the given pattern from among all possible patterns
and the empty pattern. The fundamental unit of information is the
binary decision, measured in bits. In practice, the number of decisions
can only take integer values, but if we use a continuous function,
we do not always obtain integer values; this yields the theoretical
number of decisions. For the sake of mathematical simplicity, we will
hereafter always refer to the theoretical number of decisions as the
number of decisions. For this reason, and others, determining the
amount of information is always an approximate process.
\begin{defn}
Let the information of a finite pattern $A\in X^{*}$ be the minimum
number of binary decisions necessary to uniquely specify that pattern.
Denote this by $I(A)$, where $I:X^{*}\rightarrow\mathbb{R^{\mathrm{+}}}$
is a function, and

\begin{equation}
I(A)=min\{n|A\:is\:reproducible\:by\:decision\:sequence\:(d_{1},d_{2},...,d_{n})\}\:\boxempty
\end{equation}
\end{defn}
This definition is general since it does not depend on any specific
system, purely theoretical since it explicitly incorporates all implicit
information, and is philosophically less debatable because the minimum
number of decisions represents the most general measure of information
content. At the same time, the concepts within the definition are
not precisely defined for practical or even theoretical application:
we have not fixed what exactly constitutes an elementary decision,
what reproducibility means, nor how to decompose the description of
generating a pattern into elementary decisions.

Kolmogorov complexity \cite{KOLMOGOROV} is a special case of the
above definition of information, which determines the number of decisions
required to describe patterns using a universal Turing machine:
\begin{defn}
Let $U$ be a fixed universal Turing machine. Then, the Kolmogorov
complexity $I_{K}(A)$ of a finite pattern $A\in X^{\ast}$ is defined
as:

\begin{equation}
I_{K}(A)=min\{\mid B\mid\LyXThinSpace:\LyXThinSpace U(B)=A\}
\end{equation}

where $\mid B\mid$ denotes the length of the binary program (bit
sequence), and the minimum is taken over all programs $B$ that, when
used as input to the universal Turing machine $U$, produce $A$ as
output.
\end{defn}
Unfortunately, Kolmogorov complexity is generally not computable \cite{CHAITIN_BINARY_SEQENCES}.
However, the information content can be determined very precisely
and explicitly in certain edge cases. These include numbers, constant
patterns, uniformly distributed random patterns, and patterns with
well-defined statistical properties, such as those generated by ergodic
Markov processes.

When seeking more accurate methods for defining information in both
theoretical and practical contexts, it is essential to first examine
these edge cases and the aforementioned special instances.

\section{Information Content of a Constant Pattern}

For a constant and finite pattern $A\in\{a\}^{\ast}$, no information
is required to determine the individual elements of the pattern since
it consists of the repetition of a single element. Only the length
of the pattern, $n$, carries information, which requires at most
$\lceil log_{2}n\rceil$ decisions (or bits of information) to determine,
as each decision halves the possible options. For simplicity and mathematical
tractability, we use the theoretical approximation $log_{2}n$.

Starting from the information content of integers, one can calculate
the information content of any pattern composed of identical elements
if we interpret information as the process of selecting the given
pattern from all possible sub-patterns, including the zero-length
pattern.

\begin{table}[H]
\begin{centering}
\begin{tabular}{|c|c|}
\hline 
$1$ & $()$\tabularnewline
\hline 
$2$ & $a$\tabularnewline
\hline 
$3$ & $aa$\tabularnewline
\hline 
... & ...\tabularnewline
\hline 
$n+1$ & $aaa...a$\tabularnewline
\hline 
\end{tabular}
\par\end{centering}
\caption{In the case of a constant pattern of length $n$, the definition of
information can be simplified to selecting from $n+1$ elements, where
$()$ denotes the empty pattern.}

\end{table}

The logarithm of the total number of possible patterns that can be
formed from the elements of the pattern gives the information content
of the specific pattern. In that case, the information of $A$ is:

\begin{equation}
I_{const}(A)=log_{2}(n+1)
\end{equation}

Using $n+1$ instead of $n$ is more practical because it allows the
information content of an empty pattern to be defined, acknowledging
that the empty pattern, as a possibility, also carries information.
It is easy to see that among finite patterns, constant patterns have
the lowest information content, as non-constant patterns require more
decisions due to the presence of different elements. These additional
decisions increase the information content.

We do not use the formula $log_{2}(n)$ because then subadditivity
is not satisfied for patterns of length one. The condition for subadditivity
is $I_{rand}(ab)\leq I_{rand}(a)+I_{rand}(b)$. If we used the formula
$log_{2}n$, then for the concatenation of patterns of length $n=1$,
subadditivity would not be satisfied: $log_{2}(2)\nleq log_{2}(1)+log_{2}(1)$.
In the case of the formula $log_{2}(n+1)$, however, subadditivity
is satisfied for all $n\geq0$:

\begin{table}[H]
\begin{centering}
\begin{tabular}{|c|c|c|}
\hline 
$n_{1}$ & $n_{2}$ & Subadditivity\tabularnewline
\hline 
\hline 
$0$ & $0$ & $log_{2}1\leq log_{2}1+log_{2}1$\tabularnewline
\hline 
$0$ & $1$ & $log_{2}2\leq log_{2}1+log_{2}2$\tabularnewline
\hline 
$1$ & $1$ & $log_{2}3\leq log_{2}2+log_{2}2$\tabularnewline
\hline 
$1$ & $2$ & $log_{2}4\leq log_{2}2+log_{2}3$\tabularnewline
\hline 
$2$ & $2$ & $log_{2}5\leq log_{2}3+log_{2}3$\tabularnewline
\hline 
\end{tabular}
\par\end{centering}
\caption{The fulfillment of subadditivity in the case of uniformly distributed
random sequences of different lengths. }

\end{table}

\section{Information Content of a Uniformly Distributed Random Pattern}

A finite pattern $A\in X^{*}$ with a uniform distribution can be
generated such that each element of the pattern results from an independent
decision requiring $log_{2}\left(k\right)$ bits, where $k$ is the
number of possible symbols. Considering that we can select from $n+1$
patterns of different lengths, the pattern’s information content is:

\begin{equation}
I_{rand}(A)=log_{2}\sum_{i=0}^{n}k^{i}
\end{equation}

If $k=1$, i.e., if $A$ is a constant pattern, the formula simplifies
to the constant pattern formula:

\[
I_{rand}(A)=log_{2}\sum_{i=0}^{n}1^{i}=log_{2}(n+1)
\]

The complexity of $I_{rand}\left(A\right)=log_{2}\sum_{i=0}^{n}k^{i}=log_{2}\frac{k^{n+1}-1}{k-1}$
is $O(n\cdot log_{2}(k))$, Thus, for sufficiently large $n$ and
$k$, the approximation $I_{rand}(A)\approx n\cdot log_{2}(k)$ also
holds. We do not use the formula $n\cdot log_{2}(k)$ because, in
the case of unit-length patterns, subadditivity would not hold: $2\cdot log_{2}(2)\nleq1\cdot log_{2}(1)+1\cdot log_{2}(1)$
and it would not yield accurate values for constant patterns either.

\section{Information Content of Patterns Generated by an Ergodic Markov Process}

Let $A\in X^{*}$ be a pattern that can be generated by an ergodic
Markov process, and let $f_{rel}(x_{i})=\frac{f(x_{i})}{n}$, $x_{i}\in X$,
$i=1,...,k$ be the relative frequencies of the values in the pattern.
Shannon's original formula $I_{Shannon}(A)=n\cdot\sum_{i=1}^{k}f_{rel}(x_{i})log_{2}\frac{1}{f_{rel}(xi)}$
\cite{SHANNON_MTC} would not be compatible with the formulas for
uniformly distributed patterns and constant patterns, so it needs
to be adjusted. Then, by modifying Shannon’s formula, the information
of the pattern is:

\begin{equation}
I_{mark}(A)=log_{2}\sum_{i=0}^{n}\prod_{x\in X}f_{rel}(x)^{-i\cdot f_{rel}(x)}
\end{equation}

If $k=1$, i.e., $A$ is a constant pattern, meaning $f_{rel}(x)=1$,
$x\in X$, the formula simplifies to the constant pattern formula:

\[
I_{mark}(A)=log_{2}\sum_{i=0}^{n}\prod_{x\in X}1{}^{-i}=log_{2}(n+1)
\]

For a pattern generated by a uniformly distributed process, where
$f_{rel}(x_{i})=\frac{1}{k}$, $x_{i}\in X$, $i=1,...,k$, i.e.,
all values have identical relative frequencies, the formula simplifies
to the information formula for a uniformly distributed pattern:

\[
I_{mark}(A)=log_{2}\sum_{i=0}^{n}\prod_{x\in X}\left(\frac{1}{k}\right){}^{-i\cdot\frac{1}{k}}=log_{2}\sum_{i=0}^{n}\prod_{x\in X}k{}^{i\cdot\frac{1}{k}}=log_{2}\sum_{i=0}^{n}k{}^{i}
\]

It can be seen that $lim_{n\rightarrow\infty}I_{mark}(A)=I_{Shannon}$.
Let be $c=\prod_{x\in X}f_{rel}(x)^{-f_{rel}(x)}$. If $n\rightarrow\infty$,
then $I_{mark}(A)=log_{2}\sum_{i=0}^{n}c^{i}\approx log_{2}\left(\frac{c^{n+1}-1}{c-1}\right)\approx n\cdot log_{2}c$.
It follows that $I_{mark}(A)\approx n\cdot log_{2}\prod_{x\in X}f_{rel}(x)^{-f_{rel}(x)}$,
which, based on the properties of logarithms can be written to $I_{mark}(A)\approx n\cdot\sum_{x\in X}log_{2}\left(f_{rel}(x)^{-f_{rel}(x)}\right)$,
which can be further rewritten as $I_{mark}(A)\approx n\cdot\sum_{x\in X}f_{rel}(x)\cdot log_{2}\left(\frac{1}{f_{rel}(x)}\right)=I_{Shannon}(A)$.
\begin{thm}
For finite patterns that can be generated by an ergodic Markov process,
the value of $I_{mark}(A)$ is maximized precisely when the relative
frequencies of the values are equal, i.e., $f_{rel}(x)=\frac{1}{k},\forall x\in X\:\boxempty$.
\end{thm}
The information-measuring formula can be rewritten using a logarithm
in the form $I_{mark}(A)=log_{2}\sum_{i=0}^{n}2^{-i\sum_{x\in X}f_{rel}(x)log_{2}f_{rel}(x)}$.
Because the function $-log_{2}x$ is convex, we can apply Jensen’s
inequality: $\sum_{x\in X}f_{rel}(x)log_{2}f_{rel}(x)\leq log_{2}(\sum_{x\in X}f_{rel}(x)\cdot1)$,
which simplifies to $\sum_{x\in X}f_{rel}(x)log_{2}f_{rel}(x)\leq0$.
Equality holds if and only if all $f_{rel}(x)$ values are the same,
i.e., $f_{rel}(x)=\frac{1}{k},\forall x\in X$. Therefore, the information
content of finite patterns produced by ergodic Markov processes is
exactly maximized when every value appears with the same frequency
in the pattern, and it follows that uniformly distributed random patterns
have the maximum amount of information, with information content $log_{2}\sum_{i=0}^{n}k{}^{i}$.

Shannon defined information for ergodic Markov processes \cite{SHANNON_MTC},
but it is important to note that in practice, a significant portion
of patterns cannot be generated by an ergodic Markov process, so Shannon’s
formula cannot be used to measure information and entropy in those
cases. Among all possible finite patterns, only a relatively small
subset can be generated by an ergodic Markov process. The reason is
that patterns generated by ergodic Markov processes must satisfy certain
statistical properties and transition probabilities. Kolmogorov \cite{KOLMOGOROV}
offers a more general solution than Shannon’s method. In contrast
to Shannon information, Kolmogorov complexity can be defined for every
possible finite pattern.

\section{Information Content of General Patterns}

\subsection{General Properties of Information}

From the information of specific patterns, we can infer the general
properties of information \cite{COVER_INFO_ELEMENTS}. The following
statement is easy to see:
\begin{thm}
Let $A\in X^{\ast}$ be a pattern, let $B\in X^{\ast}$a constant
pattern, $C\in X^{\ast}$ a random pattern, and $|A|=|B|=|C|$. Then
the following inequality holds: 
\begin{equation}
I_{const}(B)\leq I(A)\leq I_{K}(A)\leq I_{mark}(A)\leq I_{rand}(C)\:\boxempty
\end{equation}
\end{thm}
The information content $I_{rand}$ of a random pattern is the largest,
while the information content $I_{const}$ of a constant pattern is
the smallest. Since Kolmogorov complexity is based on Turing machines,
it cannot always provide a description of a finite pattern as concise
as might be achievable using other methods without a Turing machine:
$I_{K}$ closely approximates the information content but can be larger.
The modified Shannon information $I_{mark}$, optimized for ergodic
Markov sequences, overestimates the information content for non-ergodic
and non-Markov processes and yields higher values for less random
patterns. 
\begin{thm}
The general properties of information:
\end{thm}
\begin{enumerate}
\item \textbf{Normalization:} $log_{2}(n+1)\leq I(A)\leq log_{2}\sum_{i=0}^{n}k^{i}$,
for any $A\in X^{n}$ and $n\in\mathbb{N}^{+}$.
\item \textbf{Subadditivity:} $I(AB)\leq I(A)+I(B)$, for any $A,B\in X^{*}$.
\item \textbf{Reversibility:} $|I(A)-I(A^{R})|\leq c$, for some $c\in\mathbb{R}_{0}^{+}$,
where $A^{R}[i]=A[n-i]$, $\forall i\in\{1,...,n\}$, for $A\in X^{*}$.
\item \textbf{Monotonicity:} $I(A)\leq I(B)$, for any $A,B\in X^{*}$,
if $A$ is a subpattern of $B$.
\item \textbf{Redundancy:} $|I(A^{r})-(I(A)+log_{2}(r))|<c$, for some $c\in\mathbb{R}_{0}^{+}$,
where $A^{r}$ denotes the pattern $A$ repeated $r$ times.$\boxempty$
\end{enumerate}
The normalization property follows from the information of the constant
pattern $I_{const}(A)=log_{2}(n+1)$ and the uniformly distributed
random pattern $I_{rand}(A)=log_{2}\sum_{i=0}^{n}k^{i}$ , as well
as from Statement 1.

Subadditivity can be easily seen in the case of a constant pattern:
$log_{2}(n+m+1)\leq log_{2}(n+1)+log_{2}(m+1)$, which, when rearranged,
becomes $log_{2}(n+m+1)\leq log_{2}(n\cdot m+n+m+1)$, and this holds
in every case. For ergodic Markov processes, let $C=\prod_{x\in X}f_{rel}(x)^{f(x)}$,
then the inequality $log_{2}\sum_{i=0}^{n+m}C^{-i}\leq log_{2}\sum_{i=0}^{n}C^{-i}+log_{2}\sum_{i=0}^{m}C^{-i},$can
be rewritten as $\sum_{i=0}^{n+m}C^{-i}\leq\left(\sum_{i=0}^{n}C^{-i}\right)\left(\sum_{i=0}^{m}C^{-i}\right)$.
Rewriting the right-hand side gives $\sum_{i=0}^{n+m}C^{-i}\leq\sum_{i=0}^{n}\sum_{j=0}^{m}C^{-(i+j)}=\sum_{i=0}^{n+m}\left(\sum_{k,l;k+l=i}C^{-i}\right)$.
For every $k$, there is at least one $(k,l)$ pair that satisfies
the conditions, which means the inner sums include the term $C^{-k}$
at least once. Therefore, the inequality holds.

Reversibility means that it makes no difference from which side we
start reading the pattern---it does not affect its information content.
This is trivial, because the interpreter can easily reverse the pattern.
Monotonicity is similarly trivial for both constant patterns and those
generated by an ergodic Markov process.

Let $A^{r}=AA...A$ be a redundant pattern of length $nr$. Then $I_{const}(A^{r})=log_{2}(nr+1)=log_{2}(n+1)+log_{2}r+log_{2}\left(\frac{nr+1}{(n+1)r}\right)$.
The expression $log_{2}\left(\frac{nr+1}{(n+1)r}\right)$ approaches
$0$ as $n$ and $r$ grow, meaning it is bounded. Hence, there always
exists some $c\in\mathbb{R}_{0}$, such that $|I(A^{r})-(I(A)+log_{2}(r))|<c$.
In the case of random and ergodic Markov processes---and more generally
as well---this relationship can be seen intuitively.
\begin{defn}
Let $A\in X^{*}$ be a finite pattern. Its minimum information is
given by the function $I_{min}:\mathbb{\mathrm{X^{*}}}\rightarrow\mathbb{R^{\mathrm{+}}}$:

\[
I_{min}(A)=log_{2}(n+1)
\]

and its maximum information is given by the function $I_{max}:\mathbb{\mathrm{X^{*}}}\rightarrow\mathbb{R^{\mathrm{+}}}$:

\[
I_{max}(A)=log_{2}\sum_{i=0}^{n}k^{i}\:\text{ \ensuremath{\boxempty}}
\]
\end{defn}

\subsection{Calculating Information Based on Kolmogorov Complexity}

The general definition of information introduced in Definition 1 is
closely related to the Kolmogorov complexity specified in Definition
2 \cite{KOLMOGOROV}, Kolmogorov complexity defines the information
content of patterns on a given universal machine as the length of
the shortest binary program code that generates these patterns.

Fixing the universal machine in Kolmogorov complexity ensures that
the information of different patterns can be compared, because there
may be a constant difference between the results of various universal
machines. This difference becomes negligible for longer patterns but
can be significant for shorter ones. The relationship between information
and Kolmogorov complexity can be described by the formula $K(A)=I(A)+c$
\cite{CHAITIN_THEORY}, where $c$ is a constant characteristic of
the universal machine used to compute $K$. Since we know the minimum
information $I_{min}(A)$ precisely, we can eliminate the constant
difference and determine how to measure information based on Kolmogorov
complexity:
\begin{defn}
Let $A\in X^{*}$ be a pattern. The information of $A$ measured by
Kolmogorov complexity is defined as

\begin{equation}
I_{K}(A)=K(A)-K_{min}(A)+I_{min}(A)
\end{equation}

where $K_{min}(A)$ is the Kolmogorov complexity of a constant pattern
of length $n$:

\[
K_{min}(A)=K(\{a\}^{n})\:\boxempty
\]
\end{defn}

\subsection{Calculating Information Based on a Compression Algorithm}

However, for general patterns, it is theoretically impossible to determine
Kolmogorov complexity (i.e., the exact amount of information) precisely---only
approximations are possible. Lossless compression algorithms are the
best approach for this \cite{LI_VITANYI_KOLMOGOROV}. When a pattern
is compressed using such algorithms, it becomes almost random due
to its high information density. If we measure the information content
of the compressed pattern by assuming it is random, we obtain an approximation
of the original pattern’s information content. A characteristic feature
of compression algorithms is that the decompression algorithm and
additional data are often included in the compressed code. For smaller
patterns, this can represent a relatively large amount of extra information,
so the resulting information must be normalized.
\begin{defn}
A function $C:X^{*}\rightarrow X^{*}$is called a compression if:
\end{defn}
\begin{enumerate}
\item $C$ is injective, that is, if $A,B\in X^{\ast}$ and $C(A)=C(B)$
then $A=B$.
\item For every $A\in X^{*}$, $\mid C(A)\mid\le\mid A\mid$.
\item There is at least one $A\in X^{*}$, such that $\mid C(A)\mid<\mid A\mid$.
$\boxempty$
\end{enumerate}
For simplicity, we define the compression function so that the set
of possible values for uncompressed and compressed patterns is the
same.
\begin{thm}
If $C:X^{*}\rightarrow X^{*}$ is a compression, then $I(A)\leq C(A)$
for any $A\in X^{*}$.
\end{thm}
\begin{defn}
Let $A\in X^{n}$ be an arbitrary pattern, and let $C:X^{*}\rightarrow X^{*}$
be any compression algorithm. Then the information of $A$ measured
by the compression algorithm $C$ is given by:

\[
I_{C}(A)=\frac{I_{max}C(A)-I_{min}^{C}(A)}{I_{max}^{C}(A)-I_{min}^{C}(A)}\cdot I_{max}(A)+I_{min}(A)
\]

where

\[
I_{min}^{C}(A)=\min_{A\in X^{n}}I_{max}(C(A))
\]
\[
I_{max}^{C}(A)=\max_{A\in X^{n}}I_{max}(C(A))\:\text{ \ensuremath{\boxempty}}
\]
\end{defn}
Although determining $I_{min}^{C}$ and $I_{max}^{C}$directly from
the definition may appear cumbersome in practice, if we take into
account that compressed patterns---due to their high information
density---are almost random and can therefore be well-modeled by
a Markov process, we may use the following approximation:

\[
I_{min}^{C}(A)=I_{mark}(C(B))
\]
\[
I_{max}^{C}(A)=I_{mark}(C(D))
\]

where $B\in X^{n}$ is an arbitrary constant pattern, and $D\in X^{n}$
is any uniformly distributed random pattern.

\begin{figure}[H]
\begin{centering}
\includegraphics[width=1\textwidth]{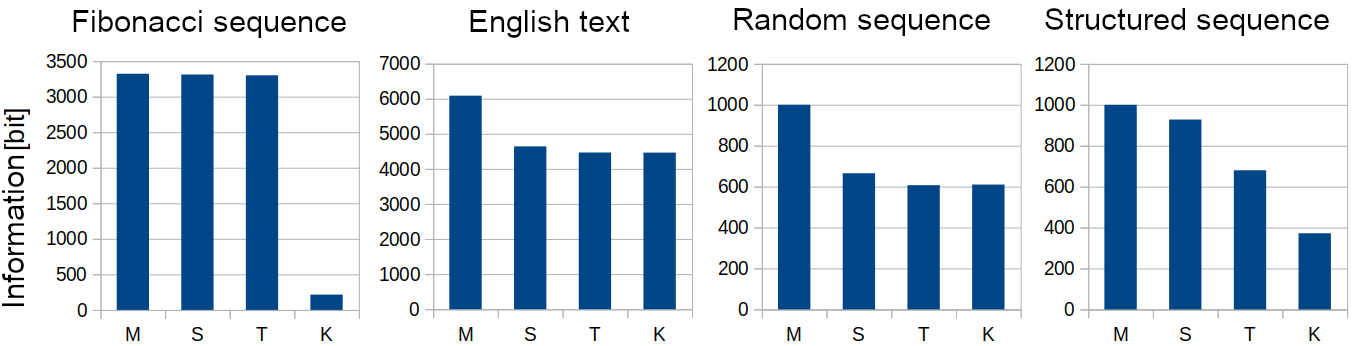}\caption{In the figure, we can see a comparison of the information values of
various 1,000-character-long patterns (APPENDIX I) that differ in
their sets of possible symbols. M denotes the maximum amount of information
possible for a pattern of the given length and symbol set. S is the
pattern’s modified Shannon information. T is the pattern’s information
as measured by the GZip compression algorithm. K is the pattern’s
approximate Kolmogorov complexity. The “random pattern” is a random
binary pattern with a certain degree of redundancy, whereas the “structured
pattern” is a 40\texttimes 25 binary character matrix in which the
'1' symbols are arranged in concentric circles.It is apparent that,
because of its seeming randomness, even the compression algorithm
could not determine the Fibonacci sequence’s information content,
whereas its Kolmogorov complexity indicated a low information content.
For the English text and the random pattern, both the Shannon-based
method and the compression algorithm provided good results. In the
case of structured text, however, the compression algorithm clearly
gives a closer approximation of the real information content than
the Shannon formula, which was originally designed for random patterns.
(The algorithms used are described in APPENDIX II--IV.)}
\par\end{centering}
\end{figure}

Different information measurement methods have varying levels of effectiveness
for different structures, so higher accuracy can be achieved by taking
the minimum of the results obtained from several measurement methods.
\begin{defn}
Let $(I_{1},I_{2},...,I_{m})$ be information measurement methods,
and let $A\in X^{n}$ be a pattern. The information of $A$, measured
using th $I_{m}$ methods, is defined as:

\begin{equation}
I_{m}(A)=\min_{i=1,...,m}I_{i}(A)\:\boxempty
\end{equation}
\end{defn}

\section{Entropy of Finite Patterns}

Unlike information, entropy is an average characteristic of a pattern,
meaning the average amount of information required to specify a single
element. In most cases, entropy is (incorrectly) identified with Shannon
entropy \cite{SHANNON_TMTC}, which only approximates the per-element
average information content well in the case of ergodic Markov processes.
Entropy calculated from Kolmogorov complexity offers a better approximation
and is more general, so it is more appropriate to define entropy based
on information content, where the method of measuring that information
is not predetermined.

If $A\in X^{*}$ is a constant pattern, $X=\{a\}$, and $()$ denotes
the empty pattern, entropy can be interpreted from a combinatorial
point of view, taking into account the empty pattern as follows:

\begin{table}[H]
\begin{centering}
\begin{tabular}{|c|c|c|}
\hline 
n & Pattern & Entropy\tabularnewline
\hline 
\hline 
$0$ & $()$ & $log_{2}(1)$\tabularnewline
\hline 
$1$ & $()(a)$ & $log_{2}(2)/2$\tabularnewline
\hline 
$2$ & $()(a)(a)$ & $log_{2}(3)/3$\tabularnewline
\hline 
\end{tabular}
\par\end{centering}
\caption{Entropy of the contant patterns.}

\end{table}

In general, entropy can be defined in this combinatorical interpretation
as follows.
\begin{defn}
Let $A\in X^{*}$ be a finite pattern. The entropy $H_{C}:X^{*}\rightarrow\mathbb{\mathbb{R^{\mathrm{+}}}}$of
the pattern is the average information content of its elements, namely:

\begin{equation}
H_{C}(A)=\frac{I(A)}{n+1}
\end{equation}

where $I$ denotes an information measurement method. $\boxempty$
\end{defn}
The factor $n+1$ in the denominator allows the formula to be interpreted
for empty patterns. For a constant pattern $A\in X^{*}$ where $|X|=1$,
we have $H_{C}(A)=\frac{log_{2}(n+1)}{n+1}$, This means that as $n$
increases, the entropy asymptotically approaches zero.

In the case of ergodic Markov processes, the entropy converges to
the Shannon entropy as $n$ increases:

\[
lim_{n\rightarrow\infty}H_{C}(A)=H(A)
\]

\begin{figure}[H]
\begin{centering}
\includegraphics[width=0.3\paperwidth]{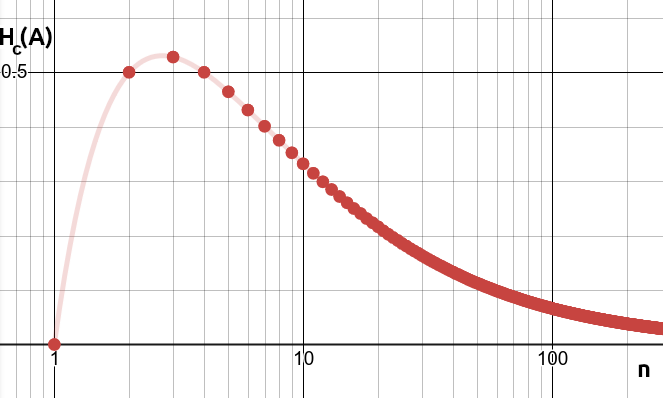}
\par\end{centering}
\caption{Entropy of a constant pattern as a function of $n$.}

\end{figure}

\section{Summary}

This paper offers a unifying view of information and entropy measures
for finite patterns that goes beyond the conventional Shannon framework.
By comparing established methods like Shannon’s entropy for ergodic
Markov processes with more general approaches such as Kolmogorov complexity,
it provides a broader perspective on measuring information content
under diverse structural conditions. Fundamental definitions for constant,
random, and Markov-generated patterns are introduced, alongside general
properties like subadditivity and redundancy. While traditional methods
frequently yield imprecise estimates for short patterns, the framework
presented here, supported by also practical, compression-based techniques,
remains robust even for very short sequences and bridges theoretical
concepts with real-world applications. This unified treatment of different
notions of entropy clarifies their suitability across various data
types, and offers mathematicians, computer scientists, and those interested
in advanced data analysis or information theory a wealth of clear
examples, formal proofs, and innovative insights into both well-known
and less-explored approaches for quantifying information in finite
sequences.

\bibliographystyle{plain}
\bibliography{References}

\appendix
\begin{center}
\textbf{\large{}\newpage APPENDIX I.}{\large\par}
\par\end{center}

The 1,000-character-long patterns used for the comparison shown in
Figure 1.

\textbf{\large{}Fibonacci sequence}{\large\par}

\texttt{\tiny{}0 1 2 3 5 8 1 3 2 1 3 4 5 5 8 9 1 4 4 2 3 3 3 7 7 6
1 0 9 8 7 1 5 9 7 2 5 8 4 4 1 8 1 6 7 6 5 1 0 9 4 6 1 7 7 1 1 2 8
6 5 7 4 6 3 6 8 7 5 0 2 5 1 2 1 3 9 3 1 9 6 4 1 8 3 1 7 8 1 1 5 1
4 2 2 9 8 3 2 0 4 0 1 3 4 6 2 6 9 2 1 7 8 3 0 9 3 5 2 4 5 7 8 5 7
0 2 8 8 7 9 2 2 7 4 6 5 1 4 9 3 0 3 5 2 2 4 1 5 7 8 1 7 3 9 0 8 8
1 6 9 6 3 2 4 5 9 8 6 1 0 2 3 3 4 1 5 5 1 6 5 5 8 0 1 4 1 2 6 7 9
1 4 2 9 6 4 3 3 4 9 4 4 3 7 7 0 1 4 0 8 7 3 3 1 1 3 4 9 0 3 1 7 0
1 8 3 6 3 1 1 9 0 3 2 9 7 1 2 1 5 0 7 3 4 8 0 7 5 2 6 9 7 6 7 7 7
8 7 4 2 0 4 9 1 2 5 8 6 2 6 9 0 2 5 2 0 3 6 5 0 1 1 0 7 4 3 2 9 5
1 2 8 0 0 9 9 5 3 3 1 6 2 9 1 1 7 3 8 6 2 6 7 5 7 1 2 7 2 1 3 9 5
8 3 8 6 2 4 4 5 2 2 5 8 5 1 4 3 3 7 1 7 3 6 5 4 3 5 2 9 6 1 6 2 5
9 1 2 8 6 7 2 9 8 7 9 9 5 6 7 2 2 0 2 6 0 4 1 1 5 4 8 0 0 8 7 5 5
9 2 0 2 5 0 4 7 3 0 7 8 1 9 6 1 4 0 5 2 7 3 9 5 3 7 8 8 1 6 5 5 7
4 7 0 3 1 9 8 4 2 1 0 6 1 0 2 0 9 8 5 7 7 2 3 1 7 1 6 7 6 8 0 1 7
7 5 6 5 2 7 7 7 7 8 9 0 0 3 5 2 8 8 4 4 9 4 5 5 7 0 2 1 2 8 5 3 7
2 7 2 3 4 6 0 2 4 8 1 4 1 1 1 7 6 6 9 0 3 0 4 6 0 9 9 4 1 9 0 3 9
2 4 9 0 7 0 9 1 3 5 3 0 8 0 6 1 5 2 1 1 7 0 1 2 9 4 9 8 4 5 4 0 1
1 8 7 9 2 6 4 8 0 6 5 1 5 5 3 3 0 4 9 3 9 3 1 3 0 4 9 6 9 5 4 4 9
2 8 6 5 7 2 1 1 1 4 8 5 0 7 7 9 7 8 0 5 0 3 4 1 6 4 5 4 6 2 2 9 0
6 7 0 7 5 5 2 7 9 3 9 7 0 0 8 8 4 7 5 7 8 9 4 4 3 9 4 3 2 3 7 9 1
4 6 4 1 4 4 7 2 3 3 4 0 2 4 6 7 6 2 2 1 2 3 4 1 6 7 2 8 3 4 8 4 6
7 6 8 5 3 7 8 8 9 0 6 2 3 7 3 1 4 3 9 0 6 6 1 3 0 5 7 9 0 7 2 1 6
1 1 5 9 1 9 9 1 9 4 8 5 3 0 9 4 7 5 5 4 9 7 1 6 0 5 0 0 6 4 3 8 1
6 3 6 7 0 8 8 2 5 9 6 9 5 4 9 6 9 1 1 1 2 2 5 8 5 4 2 0 1 9 6 1 4
0 7 2 7 4 8 9 6 7 3 6 7 9 8 9 1 6 3 7 6 3 8 6 1 2 2 5 8 1 1 0 0 0
8 7 7 7 8 3 6 6 1 0 1 9 3 1 1 7 7 9 9 7 9 4 1 6 0 0 4 7 1 4 1 8 9
2 8 8 0 0 6 7 1 9 4 3 7 0 8 1 6 1 2 0 4 6 6 0 0 4 6 6 1 0 3 7 5 5
3 0 3 0 9 7 5 4 0 1 1 3 8 0 4 7 4 6 3 4 6 4 2 9 1 2 2 0 0 1 6 0 4
1 5 1 2 1 8 7 6 7 3 8 1 9 7 4 0 2 7 4 2 1 9 8 6 8 2 2 3 1 6 7 3 1
9 4 0 4 3 4 6 3 4 9 9 0 0 9 9 9 0 5 5 1 6 8 0 7 0 8 8 5 4 8 5 8 3
2 3 0 7 2 8 3 6 2 1 1 4 3 4 8 9 8}{\tiny\par}

\textbf{\large{}English text}{\large\par}

\texttt{\tiny{}John Muir (/mj\textupsilon \textschwa r/ MURE; April
21, 1838 – December 24, 1914),{[}1{]} also known as \textquotedbl John
of the Mountains\textquotedbl{} and \textquotedbl Father of the National
Parks\textquotedbl ,{[}2{]} was a Scottish-born American{[}3{]}{[}4{]}:\LyXHairSpace 42\LyXHairSpace{}
naturalist, author, environmental philosopher, botanist, zoologist,
glaciologist, and early advocate for the preservation of wilderness
in the United States. His books, letters and essays describing his
adventures in nature, especially in the Sierra Nevada, have been read
by millions. His activism helped to preserve the Yosemite Valley and
Sequoia National Park, and his example has served as an inspiration
for the preservation of many other wilderness areas. The Sierra Club,
which he co-founded, is a prominent American conservation organization. In
his later life, Muir devoted most of his time to his wife and the
preservation of the Western forests. As part of the campaign to make
Yosemite a national park, Muir published two landmark articles on
wilderness preservation in The Century Magazine, \textquotedbl The
Treasure}{\tiny\par}

\textbf{\large{}Random sequence}{\large\par}

\texttt{\tiny{}1 1 1 1 1 0 1 1 1 1 1 1 1 0 1 1 0 1 1 0 1 1 1 1 1 1
0 0 1 1 1 1 1 1 1 1 1 1 1 0 1 1 1 1 1 1 1 1 1 1 1 1 1 1 1 1 1 1 0
0 0 0 0 1 1 1 1 0 1 1 1 1 1 1 1 1 1 1 1 1 1 1 1 1 1 1 1 1 1 1 1 1
1 1 1 0 1 0 0 0 0 0 1 1 1 1 1 1 1 1 1 1 1 1 1 0 0 0 0 0 0 0 0 1 1
1 1 1 1 1 1 1 1 0 0 0 0 0 1 1 1 1 1 1 1 0 1 1 1 1 1 1 1 1 1 1 1 1
1 0 1 1 1 1 1 0 0 1 0 1 1 1 1 1 0 0 0 0 0 1 1 1 1 1 1 0 0 1 1 1 1
1 1 1 1 1 1 1 1 1 1 1 0 1 0 1 1 1 1 1 1 1 1 0 1 1 1 1 1 1 1 1 1 1
1 1 0 1 1 1 1 1 1 1 1 0 1 1 1 1 1 1 1 1 1 1 1 1 1 1 1 1 1 1 1 1 0
0 0 0 0 0 0 0 1 1 1 1 1 1 1 1 1 1 1 0 1 1 0 1 1 0 1 1 1 1 1 1 1 1
1 1 1 1 1 1 1 1 1 1 1 1 0 1 1 1 1 1 1 1 1 1 1 1 1 1 1 0 1 1 0 1 1
1 1 1 1 1 1 1 1 1 1 0 1 1 1 1 1 1 1 1 1 1 1 1 1 1 1 1 1 1 1 1 1 1
1 0 1 0 1 1 1 1 1 1 1 1 1 1 1 1 1 1 1 1 1 1 1 1 0 1 1 0 0 0 0 0 1
1 1 1 1 1 1 1 1 1 0 1 1 0 1 1 1 1 1 1 1 1 1 1 1 1 1 1 1 0 1 1 1 1
1 1 1 1 1 1 1 1 1 1 1 1 1 1 0 1 1 1 1 1 0 1 0 0 0 0 0 1 1 1 0 1 1
0 0 1 0 1 1 1 1 1 1 1 1 0 1 1 1 1 1 1 1 1 1 1 1 1 1 1 1 0 1 1 1 1
1 1 1 0 1 1 1 1 1 1 1 1 1 1 1 1 0 1 1 1 1 0 1 1 1 1 1 1 1 1 1 1 1
1 1 1 1 1 1 1 1 1 1 1 1 1 1 1 1 1 1 1 1 1 1 1 1 1 1 1 1 1 1 1 1 1
1 1 1 1 1 1 1 1 1 1 1 1 1 1 1 1 1 1 1 1 1 1 1 1 1 1 1 1 1 1 1 1 1
1 1 1 1 1 1 1 1 1 1 1 1 1 1 1 1 1 1 1 1 1 1 1 1 1 1 1 1 1 1 1 1 1
1 1 1 1 1 1 1 1 1 1 1 1 1 1 1 1 1 1 1 1 1 1 1 1 1 1 1 1 1 1 1 1 1
1 1 1 1 1 1 1 1 1 1 1 1 1 1 1 1 1 1 1 1 1 1 1 1 1 1 1 1 1 1 1 1 1
1 1 1 1 1 1 1 1 1 1 1 1 1 1 1 1 1 1 1 1 1 1 1 1 1 1 1 1 1 1 1 1 1
0 0 1 1 1 1 1 1 1 0 1 0 1 1 1 1 0 1 1 1 1 1 1 1 0 0 0 0 0 1 1 1 1
1 1 1 1 1 1 0 1 1 0 1 1 1 0 0 0 0 0 1 0 1 1 1 1 1 1 1 1 1 1 1 1 1
1 0 1 1 1 1 0 1 0 1 1 1 1 0 1 1 1 1 1 1 1 1 1 1 1 1 1 1 1 1 1 1 1
1 1 1 1 1 1 0 0 0 0 0 1 1 1 1 1 1 1 1 1 1 1 1 1 1 1 1 1 1 1 0 1 1
1 1 1 1 1 0 1 1 1 1 0 1 1 1 1 1 1 1 0 0 0 0 0 0 0 0 1 1 1 1 1 1 1
1 1 1 1 1 1 1 0 1 0 1 1 1 1 1 0 0 0 0 0 1 0 1 1 1 1 1 1 0 1 1 0 1
1 0 0 1 1 0 0 1 1 1 1 1 1 1 0 1 1 1 0 1 1 0 1 1 1 1 1 1 1 0 1 1 1
1 1 1 1 1 1 1 1 1 1 1 1 0 1 1 1 0 1 1 1 1 1 1 1 1 1 1 1 1 1 0 1 1
0 1 1 1 1 0 0 0 0 0 0 0 1 1 1 1 1 }{\tiny\par}

\textbf{\large{}Structured sequence}{\large\par}

\includegraphics[width=8cm]{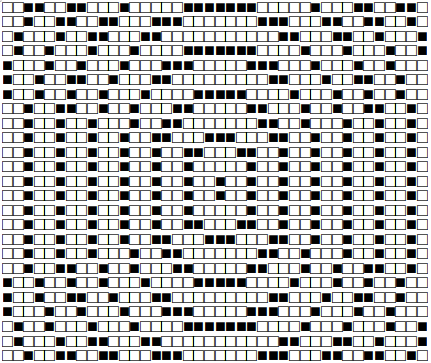}
\begin{center}
\newpage\textbf{\large{}APPENDIX II.}{\large\par}
\par\end{center}

Algorithms for Minimum and Maximum Information.

\begin{lstlisting}
public class MinInfo{

	public double minInfo(Collection values){
		if (values == null) {
			return 0;
		}
		if (values.isEmpty()) {
			return 0;
		}
		if (values.size() == 1) {
			return 1;
		}
		return Math.log(values.size() + 1) / Math.log(2);
	}

}
\end{lstlisting}

\begin{lstlisting}
public class MaxInfo{

	public double maxInfo(Collection values){ 
		if (values == null) { 
			return 0; 
		}
		if (values.isEmpty()) 
		{ 
			return 0; 
		} 
		if (values.size() == 1) {
			return 1; 
		} 
		Set atomicSet = new HashSet<>(values); 
		int k = atomicSet.size();
		int n = values.size();
		double v = n * Math.log(k) / Math.log(2);
		if (v > 500) {
			return v;
		} 
		if (k == 1) {
			return Math.log(n + 1) / Math.log(2); 
		}
		return Math.log(
			(Math.pow(k, n + 1) - 1) / (k - 1)) / Math.log(2);
	}

}
\end{lstlisting}

\newpage{}
\begin{center}
\textbf{\large{}APPENDIX III.}{\large\par}
\par\end{center}

Algorithm for the Modified Shannon Information of a Pattern.

\begin{lstlisting}
public class ModifiedShannonInfo{

	public double modifiedShannonInfo(Collection values) {
		if (values == null || values.isEmpty()) {
	    	return 0;
		}
		if (values.size() == 1) {
			return 1;
		}
		Map<Object, Double> map = new HashMap<>();      
		for (Object x : values) {
			Double frequency = map.get(x);
			if (frequency == null) {
				map.put(x, 1.0);
			} else {
				map.put(x, frequency + 1);
			}
		}
		int n = values.size();
		if (n > 100) {
			return shannonInfo.value(values);
		}
		if (map.size() == 1) {
			return Math.log(n + 1) / Math.log(2);
		}
		for (Object x : map.keySet()) {
			map.put(x, map.get(x) / n);
		}
		double info = 0;
		for (int i = 0; i < n; i++) {
			double p = 1;
			for (Object x : map.keySet()) {
				double f = map.get(x);
				p *= Math.pow(f, -i * f);
			}
			info += p;
		}
		return Math.log(info) / Math.log(2);
	}

}
\end{lstlisting}

\begin{center}
\textbf{\large{}\newpage APPENDIX IV.}{\large\par}
\par\end{center}

Algorithm for the Information of a Pattern Measured by the GZip Compression
Algorithm.

\begin{lstlisting}
public class GZipInfo{

	private final MinInfo minInfo = new MinInfo();
	private final MaxInfo maxInfo = new MaxInfo();

	public double gZipInfo(Collection values) {
		if (input == null || input.size() <= 1) {
		return 0;
		}
	
		byte[] values = ObjectUtils.serialize(input);
	
		double gzipInfo = ArrayUtils.toGZIP(values).length * 8;
	
		double min = minInfo.minInfo(input);
		double max = maxInfo.maxInfo(input);
	
		double minGzipInfo = ArrayUtils.toGZIP(
			new byte[values.length]).length * 8;
		double maxGzipInfo = ArrayUtils.toGZIP(
			generateRandomByteArray(values)).length * 8;
	
		if (originalMax == originalMin) {
			return (newMin + newMax) / 2;
		}
	
		return newMin + ((gzipInfo - minGzipInfo) 
			/ (maxGzipInfo - minGzipInfo)) * (max - min);
	}

}
\end{lstlisting}

\end{document}